\documentclass[aps,pra,twocolumn,longbibliography,superscriptaddress]{revtex4-2}
\usepackage{amsmath,amssymb,mathrsfs}
\usepackage{natbib}
\usepackage{subfigure}
\usepackage{tabularx}
\usepackage{epsfig}
\usepackage{longtable}
\usepackage{amsfonts}
\usepackage{rotating}
\usepackage{subfigure}
\usepackage{amsmath}
\usepackage{comment}
\usepackage{bbold} 
\usepackage{multirow}
\usepackage{hhline}

\usepackage{setspace}

\def\be{\begin{equation}}
\def\ee{\end{equation}}
\def\bea{\begin{eqnarray}}
\def\eea{\end{eqnarray}}

\usepackage[unicode=true,bookmarks=true,bookmarksnumbered=false,bookmarksopen=false,breaklinks=false,pdfborder={0 0 1},backref=false,colorlinks=true]{hyperref}

\hypersetup{linkcolor=magenta,urlcolor=blue,citecolor=blue,pdfstartview={FitH},hyperfootnotes=false,unicode=true}

\def\blue{\color{blue}}

\def\be{\begin{equation}}
\def\ee{\end{equation}}
\def\bea{\begin{eqnarray}}
\def\eea{\end{eqnarray}}

\begin{document}
\title{Fermion Sampling Made More Efficient}
\author{Haoran Sun$^{\blue \dagger}$}  
\affiliation{Cavendish Laboratory, University of Cambridge, Cambridge, CB3 0HE, U.K.} 
\affiliation{State Key Laboratory of Surface Physics, Institute of Nanoelectronics and Quantum Computing, and Department of Physics, Fudan University, Shanghai 200438, China}
\author{Jie Zou$^{\blue \dagger}$} 
\affiliation{State Key Laboratory of Surface Physics, Institute of Nanoelectronics and Quantum Computing, and Department of Physics, Fudan University, Shanghai 200438, China}
\author{Xiaopeng Li}  
\email{xiaopeng\_li@fudan.edu.cn}
\affiliation{State Key Laboratory of Surface Physics, Institute of Nanoelectronics and Quantum Computing, and Department of Physics, Fudan University, Shanghai 200438, China}
\affiliation{Shanghai Qi Zhi Institute, Shanghai 200030, China}
\affiliation{Shanghai Research Center for Quantum Sciences, Shanghai 201315, China}

\begin{abstract}
{
 Fermion sampling is to generate probability distribution of a many-body Slater-determinant wavefunction, which is termed ``determinantal point process" in statistical analysis. For its inherently-embedded Pauli exclusion principle, its application reaches beyond simulating fermionic quantum many-body physics to constructing machine learning models  for diversified datasets.  Here we propose a fermion sampling algorithm, which has a polynomial time-complexity---quadratic in the fermion number and linear in the system size.  This algorithm is about $100\%$ more efficient in computation time than the best known algorithms. In sampling the corresponding marginal distribution, our algorithm has a more drastic improvement, achieving a scaling advantage.  We demonstrate its power on several test applications, including sampling fermions in a many-body system and a machine learning task of text summarization, and  confirm its improved computation efficiency over other methods  by counting floating-point operations. 
}

\end{abstract}

\date{\today}

\maketitle

The fermion sampling problem is to generate a position configuration of a quantum system composed of  $N$ fermions, ${\bf x} = (x_1, x_2, \ldots, x_N)$, according to Slater-determinants, 
where the position $x_j$ is an integer running from $1$ to the system size $L$. The single-particle wavefunctions of the fermions are stored in a $L\times N$ matrix $U$ 
as a series of column vectors. The orthonormal condition of the wavefunctions implies $U^\dag U = \mathbb{1}$. In this sampling problem, the  configuration ${\bf x}$ is drawn from a probability distribution 
\be 
P({\bf x}) = \frac{1}{N!} | {\rm det} (U_{{\bf x},{\bf n}} )|^2 , 
\label{eq:sampletask} 
\ee 
with ${\bf n } = (1, 2, \ldots, N)$. This problem arises in modeling fermionic quantum many-body physics, whose computation complexity causes numerical challenges  in simulating many-electron interference~\cite{li2019} and variational Monte Carlo calculations~\cite{becca_sorella_2017}. Such numerical simulations play important roles in understanding strongly correlated quantum many-body physics including electron correlation effects in molecules~\cite{1999_Kohn_RMP} and in quantum materials~\cite{2017_Moore_NatPhys}, and equation-of-state of neutron stars~\cite{1999_Baker_PRC}. A standard traditional approach to perform fermion sampling  in physics is through Markov chain  methods, where the computation cost of the direct calculation of Slater-determinants is avoided by taking sequential local updates. 
The resultant sampling complexity is reduced to ${\cal O} (N^3)$.
However this approach becomes inefficient in presence of large autocorrelation length in the Markov chain, which generically appears in the simulation of quantum criticality~\cite{2005_Coleman_Nature} and many-body localization in disorder systems~\cite{2006_BAA}. 

In statistical analysis and machine learning applications, the fermion sampling problem has been termed ``determinantal point process" (DPP)~\cite{macchi1975coincidence}. 
It has deep connection with random matrix theory~\cite{RMT_tao2012} and completely integrable systems~\cite{IntegrableDPP_jimbo1980} in mathematics. 
Due to the built-in negative correlation by the quantum Pauli exclusion principle, DPP has been adopted as an elegant way to enhance diversity~\cite{ReviewDPP_kulesza2012}. 
This has triggered tremendous interests in machine learning, as the ingredient of diversity is fundamentally crucial in such learning tasks as recommendation systems~\cite{RecommmandDPP_chen2018}, text summarization, and image searching~\cite{ImageDPP_kulesza2011k}, and is difficult to characterise with other models~\cite{ReviewDPP_kulesza2012}. 
One basis underlying these machine learning applications is the Hough-Krishnapur-Peres-Virag (HKPV) algorithm \cite{hough2006determinantal}, 
whose sampling complexity has been improved from the original ${\cal O}(N^3L)$ to ${\cal O} (N^2 L)$ with certain modification~\cite{ModifiedHKPV_Gil2014}.

Here, we propose a novel fast fermion sampling (FFS) algorithm with a time complexity of ${\cal O} (N^3 + N^2 L)$. This is an explicit sampling algorithm, free of the autocorrelation problem present in the Markov chain sampling, and about $100\%$ more efficient than the modified HKPV algorithm in computation time. 
The ${\cal O} (N^3)$ tail in the time complexity is less dominant in general since the system size $L$ has to be  larger than the particle number $N$ due to Pauli exclusion principle.
The improvement is more dramatic for sampling the corresponding marginal distribution---our algorithm then has a scaling advantage in the computation time. 
We demonstrate our algorithm in both quantum physics and machine learning examples. In application to sampling free fermions in a double-well, and interacting fermions in disorder potentials, we find the sampling error by our FFS algorithm is much smaller (more than ten times smaller in certain cases) than the Markov chain sampling given the same level of computation resources. Their distinction coincides with the autocorrelation length, as FFS is an explicit sampling algorithm. 
In application to text summarization, we find our FFS algorithm outperforms the modified HKPV algorithm in computation efficiency---the float point operations with FFS are $50\%$ smaller than the modified HKPV for the same learning task. 
This $100\%$ improvement in the computation efficiency remains with increasing text-summary length, meaning our FFS algorithm is substantially more powerful for heavy text summarization tasks.

Our fast fermion sampling algorithm is based on a mathematical formula established for the Slater determinant wavefunction~\cite{li2019}, 
\be 
 \sum_{\bf m}   \left[
  \prod_k P(x_k| x_1, \ldots x_{k-1}; {\bf m} ) \right] = |{\rm det} (U_{{\bf x}, {\bf n}})|^2, 
  \label{eq:mathbas} 
\ee 
with ${\bf m } = (m_1, m_2, \ldots m_N) $ running over all permutation of ${\bf n}$, and  the conditional probability distribution, 
$
P(x_{k} \mid x_{1}, \ldots x_{k-1}; {\bf m} ) = \frac{1}{k !}\left|{\rm det}\left[ U_{(x_{1} \ldots x_{k}), (m_1 \ldots m_k)}\right]\right|^{2}. 
$
We first generate a vector ${\bf m} $ as a random permutation of ${\bf n}$. Then the fermion position $x_k$ is drawn according to the conditional probability distribution, with  
 the index $k$ iteratively  increased from $1$ to $N$ step by step. It is guaranteed by Eq.~\eqref{eq:mathbas} that the sampled configuration ${\bf x} = (x_1, x_2, \ldots x_N)$ obeys the required distribution in Eq.~\eqref{eq:sampletask}.

One key observation we make here is that the sampling $x_k$ from the conditional probability $P(x_{k} \mid x_{1}, \ldots x_{k-1}; {\bf m} )$ does not require calculation of determinants, for their geometrical interpretation. A determinant is equal to the volume of the high dimensional parallelotope spanned by the row vectors. Denoting the $k$-dimensional row vectors in $U_{(x_{1} \ldots x_{k}), (m_1 \ldots m_k)}$ as \{${\bf u}_{x_1}$, ${\bf u}_{x_2}$, \ldots ${\bf u}_{x_k}$\}, the geometrical interpretation of the determinant implies an important property for  the conditional probability, 
\be 
P(x_{k} \mid x_{1}, \ldots x_{k-1}; {\bf m} ) \propto |{\bf u}_{x_k} ^T {\bf h} ( {\bf u}_{x_1}, {\bf u}_{x_2}, \ldots, {\bf u}_{x_{k-1}})|^2,  
\ee 
with ${\bf h}$ a $k$-dimensional normal vector perpendicular to the $k-1$ vectors \{${\bf u}_{x_1}, {\bf u}_{x_2}, \ldots, {\bf u}_{x_{k-1}}$\}. 
The normal vector can be efficiently calculated with an iterative Gaussian elimination method, which consumes $2kN$ operations ({\it Supplementary Information}).  
Calculating the inner product ${\bf u}_{x_k}^T {\bf h}$ for all $x_{k}$ yields a computation cost of $2Lk$.  We then have 
a sampling complexity at $k$-th step, with $2Nk+2Lk$ operations. 
Running $k$ from $1$ to $N$, the overall operations of generating one sample configuration $\bf{x}$ is $LN^2+N^3$, to the leading order of $L$ and $N$. However, the modified HPKV algorithm requires $2LN^2$ operations in total~\cite{ModifiedHKPV_Gil2014}, which is nearly the twice of our method in the limit of  $L\gg N$.
Furthermore, different sample configurations are completely uncorrelated, in sharp contrast to the Markov chain sampling.
In particular, for a computation task of sampling the marginal distribution of $N_<$ ($<N$) fermions, the computation cost of our FFS algorithm is ${\cal O} (LN_<^2+N_<^3)$, having a scaling advantage over the modified HKPV whose computation cost goes as ${\cal O} (LNN_<)$.

\begin{figure}[htp]
\vspace{.2cm}
\centerline{\includegraphics[angle=0,width=.5\textwidth]{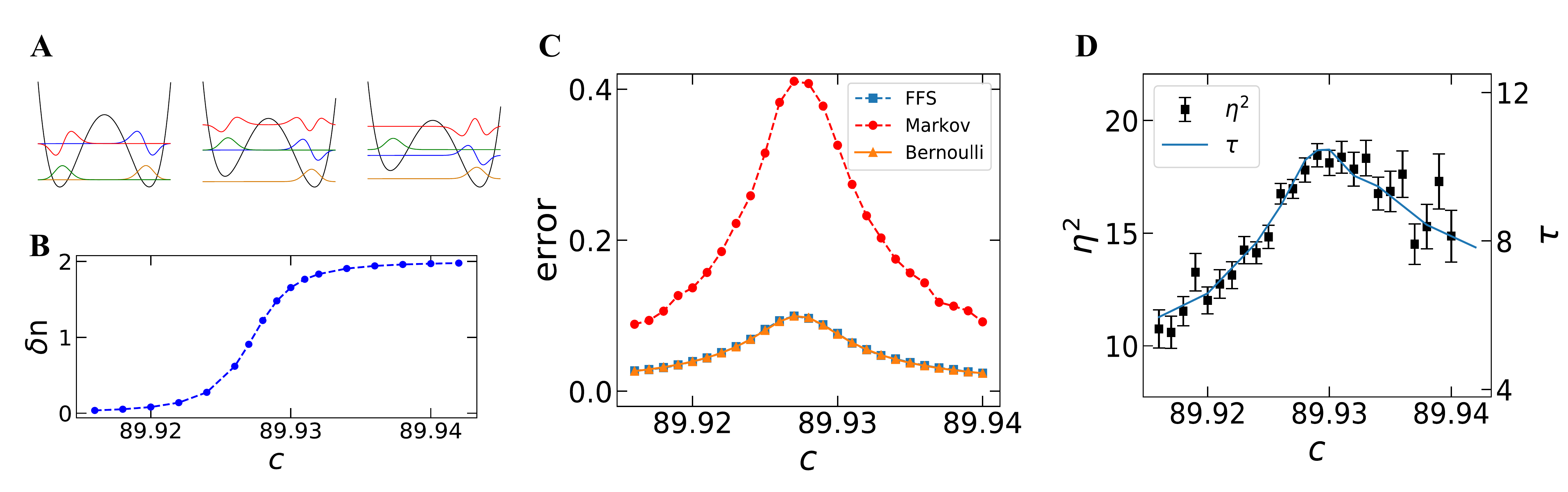}}
\caption{Stimulation of double well potential via FFS and Markov chain sampling methods. 
(A) A schematic illustration of double well potential and its wave functions in three stages: balanced, transition, and imbalanced. 
Colorful lines  illustrate different orbital wave functions and the relative positions correspond to the energy levels. 
All the orbits are strongly localized in one potential well except for the highest orbit during the transition stage. 
(B) Particle number density difference in two potential wells under different level of skewness. 
(C) Error of FFS and Markov chain methods around the transition point, comparing to a  theoretical error assuming a Bernoulli distribution (Supplementary Information). 
(D) illustrates that squared error ratio $\eta$ is strongly related to autocorrelation length $\tau$ in the Markov chain method.}
\label{figure_DW}
\end{figure}

In order to benchmark our algorithm, we first consider a test case of sampling fermions in a double-well potential, and compare our algorithm with the  Markov chain sampling. 
We take a standard form of a one-dimensional double-well potential, $V(x) = a(x ^2-b)^2-cx $, with $x$ the spatial coordinate. The system Hamiltonian upon space discretization is given by
$$H = -t \sum_{\langle ij \rangle} c_{i}^{\dagger} c_{j} + \sum_i v_i n_i,$$
with parameter $v_i$ taken from the double well potential. 
The parameters $a$, and $b$ in  $V(x)$ controls the depth of potential well and the parameter $c$ controls its skewness. 
In order to achieve strong particle localization of the system, the depth parameter $a$ is set $10^3$ times of the tunneling strength in our implementation.
In small $c$ limit, fermionic particles are evenly distributed in both potential well at their ground state. 
A finite value of $c$ introduce a break of symmetry to the system, and induce a balance-imbalance transition. 
At a critical $c$ value, the difference between the particle number in two potential wells exhibits a sharp transition from zero to two, and an increase in autocorrelation length is observed. 
We measure the particle density in each potential well, and compare the performance of the Markov chain and FFS algorithms. 
Given a constant total running time, a significantly larger error is observed at the critical point for the Markov chain sampling. 
In contrast, our algorithm shows no error increase other than the physical increase of the intrinsic observable variance in the system (Fig.~\ref{figure_DW}C). 
By measuring an error ratio $\eta = \frac{\epsilon_{\text{Markov}}}{\epsilon_{\text{FFS}}}$, we show our approach significantly outperforms traditional method by achieving higher accuracy with same computational requirement (see Fig.~\ref{figure_DW}(D)).

This rise of the simulation error at phase boundary due to critical slowing down, 
is a key feature generically hindering the application of Markov chain Monte Carlo sampling methods to various models at the critical point.  
Our proposing fast fermion sampling algorithm is completely immune to the critical slowing down problem. 


\begin{figure}[htp]
\centerline{\includegraphics[angle=0,width=.5\textwidth]{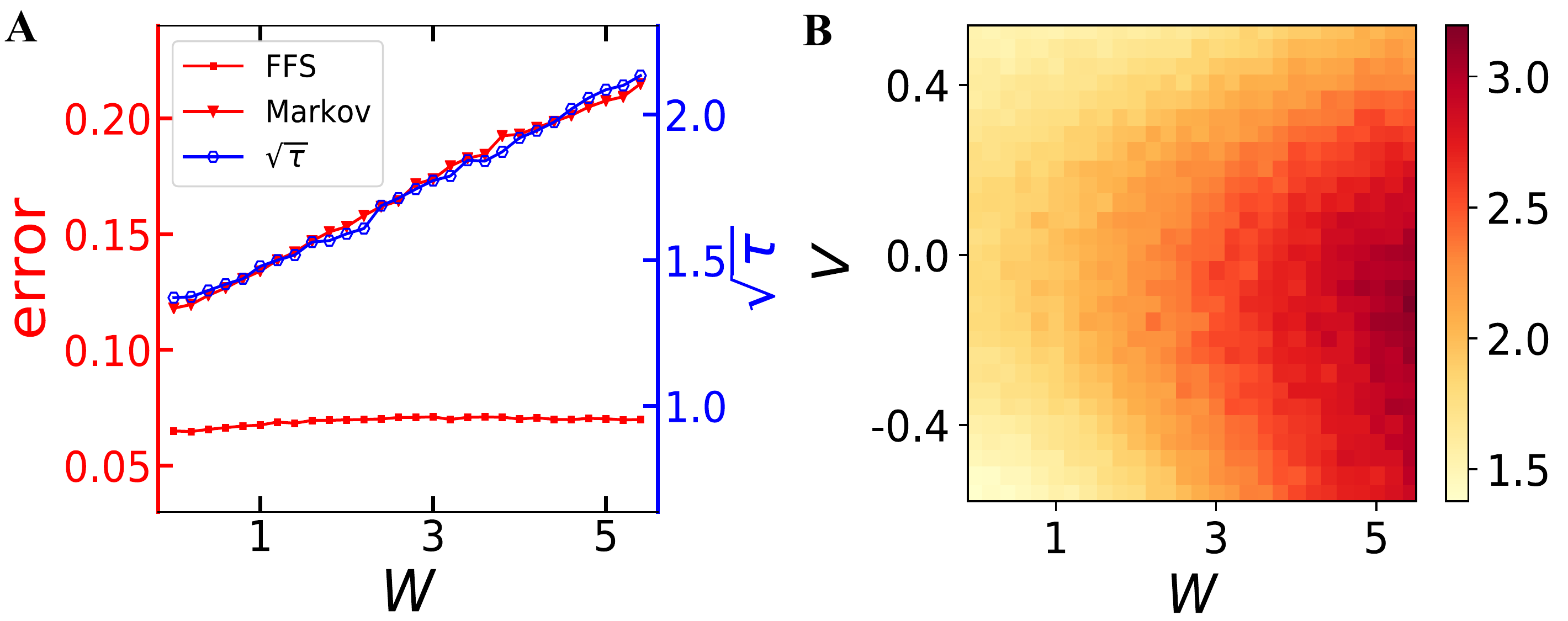}}
\caption{
Simulation of Anderson-Hubbard model via FFS and Markov chain methods.
The error is measured according to their sampling results on the occupation number on the first lattice site. 
(A) shows the errors with both methods and the square  root of autocorrelation length in the Markov chain sampling for the Anderson model ($V=0$) with increasing disorder strength $W$. 
(B), The error ratio $\eta$ between Markov chain sampling and  FFS in simulating the Anderson-Hubbard model with different interaction ($V$) and disorder strengths ($W$). 
Our algorithm outperforms for all interaction $V$ and disorder strength $W$ we choose, and the comparison is more substantial  for stronger disorder and weaker interaction.
All the calculations above are implemented on a $4 \times 4 $ square lattice with the particle number $N=8$.
}
\label{fig:Anderson}
\end{figure}

We also apply our algorithm to sample a correlated system of interacting fermions, described by a Slater-Jastrow wave function
\be
|\psi_J \rangle 
= \exp(-\frac{1}{2} V †\sum_{i} n_{i,\uparrow} n_{i,\downarrow}) |\phi _0 \rangle,
\label{eq:Jastrow}
\ee
We choose the the non-interacting part $|\phi_0\rangle$ to be $N$-fermion ground state of a two-dimensional Anderson model, 
\be  
H = -t \sum_{\langle ij \rangle} \sum_{\sigma} c_{i, \sigma}^{\dagger} c_{j, \sigma} ++ \sum_i w_i n_i,
\label{eq:Anderson}, 
\ee
with $w_i$ drawn from a uniform distribution on $[-W, W]$. 
This problem arises in variational Monte Carlo calculations of Anderson-Hubbard model.  
The occupation number on the site with index $(i=0,j=0)$ is calculated with both of our algorithm and Markov chain sampling for comparison (see Fig.~\ref{fig:Anderson}). 
In Markov chain sampling, to let the dynamics equilibrate we perform $2000$ steps of local updates first and then start to  take measurements.
For a fair comparison, we let the number of sampling configurations in our algorithm and the number of measurements in Markov chain be equal (set to be $100$ here). 
The sampling error with respect to the exact values by the two methods is shown in Fig.~\ref{fig:Anderson}---the exact values are obtained by sampling $10^6$ times with our FFS algorithm. 
The sampling error by Markov chain systematically increases with the disorder strengths, whereas the error remains at the same level for different disorder strengths. We attribute this to the localization physics or the tendency towards  localization. With larger disorder strength, the system develops tendency towards localization, and the Markov chain sampling becomes less efficient in exploring the configuration space for the localization causes nonergodicity, rendering a larger autocorrelation length.  This problem is completely absent in our algorithm for it produces independent samples. As we increase correlation effects, the distinction between our algorithm and Markov chain becomes less dramatic, which is as expected because interactions effects in general compete with disorder-induced localization~\cite{2006_BAA}. From these results, it is evident that our fast fermion sampling algorithm systematically outperforms the Markov chain method, especially when autocorrelation is significant.

\begin{figure*}[htp]
\centerline{\includegraphics[angle=0,width=\textwidth]{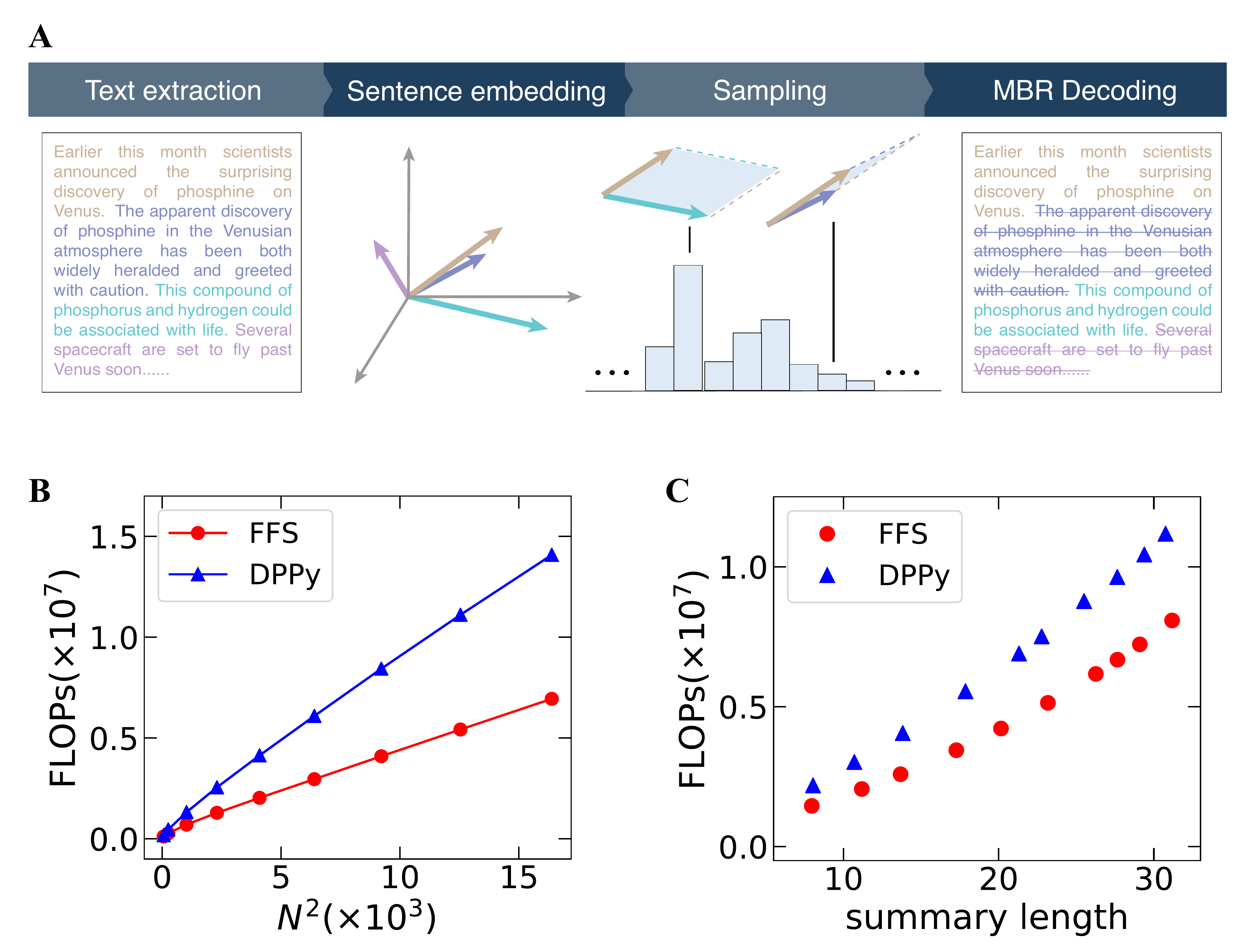}}
\caption{
Application of FFS to text summarization task. 
(A) A diagrammatic representation of the general approach of text summarization task using determinantal process. Different colors mark different sentences.
(B) A simple fermion sampling task for a direct comparison of computation time cost between FFS and DPPy algorithms. We randomly generate a $N$-particle fermionic wave function on a $L$-site 1D chain and measure the floating point operations (FLOPs) required for producing a sample. Here we choose $L=1024$, which is sufficiently larger than all possible $N$. The slope of DPPy is nearly twice that of FFS.
(C) FLOPs required to generate a summary from a 100 sentences article, in the fermion sampling step, with different required summary length.
}
\label{fig:textsum_new}
\end{figure*}

Last, we also apply the method to text summarization, to demonstrate the potential impact of the fast fermion sampling algorithm to machine learning applications. 
Text summarization is the task of  producing a concise summary of extended texts. 
One approach used in machine learning is to select sentences in the texts that reflect the most important information. 
A balance should be maintained between diversity and quality of chosen information in order to make a concise summary.
For such machine learning tasks, the quantum Pauli exclusion principle built in the fermion sampling problem, termed determinantal point process in statistical analysis,  provides an elegant approach to impose sampling repulsion~\cite{TextSum_kulesza2011}.

For demonstration, we adopt a standard approach of using DPP for text summarization~\cite{ReviewDPP_kulesza2012}, which mainly contains the following steps. First, all sentences are converted to unit column-vectors ${\bf b}_1$, ${\bf b}_2$, ..., ${\bf b}_L$---the subscript labels correspond to different sentences, with a standard word embedding algorithm~\cite{rehurek_lrec}, where the correlation among the sentences is encoded into inner products of those unit vectors.  The magnitude of each vector is then assigned to reflect the quality~\cite{ReviewDPP_kulesza2012} of the corresponding sentence as a summary, giving new vectors $\Tilde{{\bf b}}_1$, $\Tilde{{\bf b}}_2$, ..., $\Tilde{{\bf b}}_L$. The features considered here include sentence length, sentence position, personal pronouns, etc. The feature to magnitude mapping is achieved by a pretrained log-linear model ({\it Supplementary Information}). A correlation ${\bf C}$ matrix is stored with elements ${\bf C}_{ij} = \Tilde{{\bf b}}_i ^T \Tilde{{\bf b}}_j$. Its spectral decomposition is obtained as $C = \sum_j \lambda_j {\bf u}_j {\bf u}_j ^T $, with ${\bf u}_j$ the eigenvectors of dimension $L$. 
Second, we draw a sample $(x_1, x_2, \ldots x_N)$ following the standard DPP taking the matrix $C$ as a kernel. Each eigenvector ${\bf u}_j$ is selected with probability $\frac{\lambda_j}{\lambda_j+1}$, representing its importance in the correlation matrix. The selected set of vectors, 
\{${\bf u}_{j_1}$, ${\bf u}_{j_2}$, \ldots ${\bf u}_{j_N}$\}, form the matrix $U$ in Eq.~\eqref{eq:sampletask}, according to which we perform the fermion sampling with fermion particle number $N$, and system size $L$. Our fast fermion sampling algorithm is benchmarked against the modified HKPV algorithm widely used in DPP context. 
The sampled configuration $(x_1, x_2, \ldots x_N)$ marks the sentences to be kept in the summary. 
Third, as the DPP process occasionally produces low-quality summary~\cite{ReviewDPP_kulesza2012}, a most appropriate summary is produced by taking the minimum Bayes risk (MBR) decoding~\cite{GOEL2000115} ({\it Supplementary Information}).

The overall complexity of the above scheme is ${\cal O}(L^2 + MN^2L + M^2N)$, with $M$ the number of samples taken in the second step. Our algorithm is expected  to have a constant speedup compared to the modified HPKV method in the sampling procedure.
Before implementing the text summarization task, we carry out a pure fermion sampling task for a clean comparison between two sampling algorithms. 
The results are shown in Fig.~\ref{fig:textsum_new}. 
The results of the modified HPKV algorithm are obtained by the sampling module in the DPPy~\cite{DPPy_GPBV2019} open source library, which implements the modified HKPV algorithm. 
As shown in Fig.~\ref{fig:textsum_new}B, the computation cost of the two algorithms at large system size scales the same way in the floating point operations (FLOPs). The computation cost of  our FFS algorithm is about one half of the modified HKPV by counting FLOPs. This $100\%$ improvement in the computation efficiency remains as we increase the system size.  
Then we perform a test on a public CNN/Daily Mail news data set~\cite{see2017point}. The results are presented in Fig.~\ref{fig:textsum_new}C. 
We show the counting of FLOPs solely in the fermion sampling step for  a direct comparison of our FFS with the modified HKPV algorithm in the text summarization task. 
The FFS algorithm outperforms the modified HKPV  for all summary lengths consistently. 
The improvement is now  less than $100\%$, which can be attributed to the ${\cal O}(N^3)$ tail in the computation cost of FFS. 
But still the practical improvement with our FFS algorithm is quite substantial at large summer length. 
Our approach then has promising applications in accelerating the machine learning task of multiple-text summarization, 
where the summary length can be a large number to cover all the important information of multiple texts.

\paragraph*{Discussion.} 
We expect  the developed fast fermion sampling algorithm to have  wide applications in machine learning and quantum many-body physics. In all machine learning tasks where describing data repulsion is important, DPP is an elegant probabilistic model, and the fast fermion sampling algorithm has potential to reduce the computation cost substantially. In applications to simulating strongly interacting fermions or solving quantum chemistry problems, the fast fermion sampling algorithm may be integrated with other importance sampling methods to reach optimal computation performance. 
 
We remark here that in performing fermion sampling for continuous models, a direct implementation of our fast fermion sampling algorithm could be  inefficient because the time cost grows with $L$, which is typically a large number in discretizing a continuous model to minimize discretization error. In Supplementary Information, we show how this algorithm adapts to sampling continuous models with minor modification, which leads to  a computation cost of ${\cal O} (N^3)$.

\paragraph*{Acknowledgement.}
We acknowledge helpful discussion with Yang Qi. 
This work is supported by National Program on Key Basic Research Project of China (Grant No. 2017YFA0304204), National Natural Science Foundation of China (Grants No. 11934002, and 11774067), Shanghai Municipal Science and Technology Major Project (Grant No. 2019SHZDZCX01). We also acknowledge the support by the Talented Student Program on the Fundamental  Disciplines from Ministry of Education of China.

$^{\blue \dagger}$ These authors contributed equally to this work. 

\bibliography{scibib}

\newpage

\begin{widetext} 
\renewcommand{\theequation}{S\arabic{equation}}
\renewcommand{\thesection}{S-\arabic{section}}
\renewcommand{\thefigure}{S\arabic{figure}}
\renewcommand{\thetable}{S\arabic{table}}
\setcounter{equation}{0}
\setcounter{figure}{0}
\setcounter{table}{0}

\newpage

\begin{center} 
{\Huge \bf Supplementary Information} \\
\end{center} 


\section{The geometrical approach for determinantal sampling} 
\label{sec:caldet} 
As described in the main text, the major computation cost of our fast fermion sampling algorithm consists in the calculation of the determinants in the conditional probability distribution in Eq. (2). 
In sampling $x_k$ with $k$ running from $1$ to $N$, we need to calculate $L-k+1$ number of determinants [${\rm det } \, U_{(x_1, \ldots x_k), (m_1, \ldots m_k)}$]  in each $k$-step, whose direct computation is too costly.

In this section, we show how these determinants are calculated efficiently. Here we exploit the geometric interpretation of determinants, that is, the absolute value of the determinant ${\rm det } \, U_{(x_1, \ldots x_k), (m_1, \ldots m_k)}$ equals to the volume spanned by its $k$-dimensional row vectors \{${\bf u}_{x_1}$, ${\bf u}_{x_2}$, \ldots ${\bf u}_{x_k}$\}. 
In the $k$-th step, since the first $k-1$ row vectors of the determinants are fixed, we can take them as the base and the remaining task of computing the volume is to get the height ({\it see main text}).

To this end, we can calculate the unit normal vector $\bf{h}$ first, and then the height equals to the projection of the last row vector $\bf{u}_{x_k}\equiv  U_{x_k, (m_1, \ldots m_k)}$ onto $\bf{h}$. 
Of course this normal vector can be derived by implementing the standard Gaussian elimination method on the $(k-1) \times k$ submatrix $U_{(x_1, \ldots x_{k-1}), (m_1, \ldots m_k)}$, which takes a computation time of $O(k^3)$. However, instead of a naive implementation, we can compute the normal vector iteratively, since Gaussian elimination operations on the first $k-2$ rows of $U_{(x_1, \ldots x_{k-1}), (m_1, \ldots m_k)}$ at step $k$ are exactly the same as those of $U_{(x_1, \ldots x_{k-2}), (m_1, \ldots m_{k-1})}$ at step $k-1$. 
Therefore, at step $k$, only the first $k-2$ entries in the $(k-1)$-th row vector $\bf{u}_{x_{k-1}}$ needs to be eliminated.
But we have not dealt with the $k$-th column in $U_{(x_1, \ldots x_{k-1}), (m_1, \ldots m_k)}$ by now, which is also appended at the beginning of step $k$. This problem can be naturally solved if we always carry out Gaussian elimination procedures in the full $N$-dimensional space, that is, entries in column $m_{k+1}, m_{k+2}, ..., m_{N}$ are also updated at step $k$. 

With the setup above, we can give a quantitative description for the complexity of our algorithm. It requires $\sum_{j=0}^{k-3} (N-j) \approx Nk$ multiplications and approximately the same number of additions to implement the iterative Gaussian elimination method at the $k$-th step. Then the normal vector $\bf{h}$ can be determined with another $ k(k-1)/2 $ additions and summations. Taking into account the computation of inner products of $L-k+1$ candidate vectors for $\bf{u}_{x_k}$ with the normal vector $\bf{h}$, our algorithm consumes $2Nk+2Lk$ operations (including additions and summations) at step $k$, to the leading order of $N$ and $L$. Therefore, with $k$ running from 1 to $N$, the total operation count of the whole algorithm is $LN^2+N^3$.

\section{Determination of error and autocorrelation length}
The error is defined by the root mean square of absolute error from single trial. 
More specifically, for both algorithms, we take 100 samples in each trial and obtain an averaged observable value from these samples. 
The trial is repeated for approximately $10^4$ times (vary for different models) to accumulate sufficient data. 
The absolute error for each trial is taken to be the difference between value from single trial and an average of all trials 
in the simulation of Anderson-Hubbard model.
But we replace the average of all trials with the theoretical value in the double-well model because here we only deal with the noninteracting case which is exactly solvable.
The presented error is a root mean square of all these errors from different trials. 

The autocorrelation length (ACL) is calculated naively from a logarithm fitting of the autocorrelation function $\hat c$ (ACF). Let $\left\{f_{n}\right\}_{n=1}^{N}$ be a finite chain of input data with mean $\mu$, the ACF is defined by
\begin{equation}
\hat{c}(i)=\frac{1}{N-i} \sum_{n=1}^{N-i}\left(f_{n}-\mu\right)\left(f_{n+i}-\mu\right)
\end{equation}
The ACF in a general Monte Carlo process is modeled by a exponential decrease $\hat c(i) = \exp({-\frac{i}{\tau})}$. 
A logarithm fitting give raise to an unbiased estimation of the ACL. 
We can also compute the ACL by discretize the exponential relation above, that is to say,
\begin{equation}
\tau= \sum_{i=0}^{\infty} \frac{\hat c(i)}{\hat c(0)}.
\end{equation}
In practice we choose a sufficiently large cutoff for $i$ instead of a summation to infinity.

\section{Simulation details on Double Well}
The system is stimulated in 64-site 1D lattices, with four non-interacting spinless fermions. 
The double well potential is defined by 
\begin{equation}
    V(x_i) = a(x_i^2-b)^2-cx_i
\end{equation}
with $a = 4096$, $b = 2$ and a varying $c$. The position of each lattice, $x_i$, ranges evenly from $-2$ to $2$. The system tunneling strength, $t$, was taken to be $\frac{\pi^2}{6}$. These values are chosen for particles to stay strongly localized in certain potential well. A sharp transition between balance and imbalance states of the system was observed around $c = 89.929$. 

The error of both methods is measured by comparing an exact solution and an average of 100 consecutive samples. They are consequently compared with a theoretical estimation of the natural error aroused from inherent uncertainty. By assuming an ideal sampling of the system be Bernoulli process, with particles falls into either side of the potential well, the theoretical error is expressed by
\begin{equation}
    \epsilon = \sqrt{\frac{p(1-p)}{M}}
\end{equation}
where $p = \sum_{x_i<0} |\phi_i|^2$ be the probability of the particle with highest energy falls into the negative side of the potential, and $M$ be the number of samples taken.

\section{Simulation details on Anderson-Hubbard model}
In the main text we show the strong relevance between error $\epsilon$ and autocorrelation length $\tau$ in variational Monte Carlo (VMC) method through the collapse of the curves of $\epsilon$ and $\sqrt{\tau}$. 
This square relation is chosen from a data fitting analysis, as shown in Figure S2{\bf A}. 
This relation is also reasonable from a perspective of the central limit theorem where error is proportional to $1/\sqrt{N}$ and $N$ is the amount of statistically independent samples. 
The autocorrelation length reduces the effective sample number up to a factor $1/\tau$ and the above relation is thus derived.

For the interacting Anderson-Hubbard model, we have calculated the error ratio of Markov chain sampling and FFS methods within a broad parameter regime. 
In in Figure S2{\bf B}. we show that for interacting cases the error of FFS still nearly remains a constant and the one with Markov chain sampling grows linearly with the increasing disorder strength. 
Besides, the appearance of interaction has nothing to do with the error of FFS but relieve the trouble brought from the localization for Markov chain sampling. 
Again, we compare the error and the square root of autocorrelation length in Figure S2{\bf C}, and the relevance between them is tested. 
All the facts above are consistent with the conclusions in our main text.

\begin{figure}[htp]
\centerline{\includegraphics[angle=0,width=\textwidth]{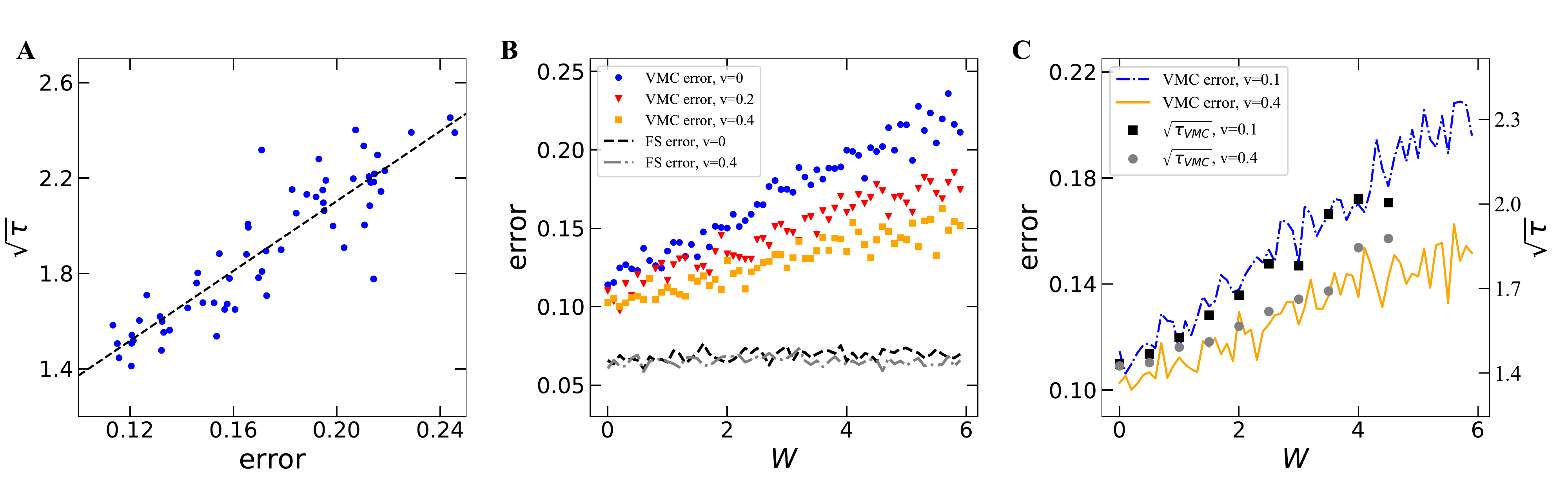}}
\caption{Simulation details on Anderson-Hubbard model.  
{\bf (A)}, a linear fit between error and the square root of autocorrelation length $\sqrt{\tau}$. 
{\bf (B)}, shows the error of FFS and Markov chain sampling with increasing disorder strength $W$ for both noninteracting and interacting case.
{\bf (C)}, shows the error and autocorrelation length in Markov chain sampling with increasing disorder strength for different interacting strengths.
All the simulations above are implemented on a $4\times 4$ square lattice and the particle number is 8.
The error is measured according to the occupation number of the first site.
}
\label{figs:suppAH}
\end{figure}

\section{Implementation of our method on the Slater-Jastrow wave function}
Jastrow wavefunction is a commonly used trial wave function in VMC, which is written as
\be
|\psi_J \rangle = J_v |\phi_0 \rangle = \exp(-\frac{1}{2} \sum_{i,j} v_{ij} n_i n_j) |\phi _0 \rangle, 
\ee
where $n_i$ labels the particle number on site i and $|\phi _0 \rangle$ is a non-interacting wave function.
$v_{ij}$s are variational parameters to be optimized and the Jastrow factor $\hat{J}_v$ is expected to give a proper description on the strong correlation between particles.
In the main text we choose the Gutzwiller factor, a simple version of Jastrow factor with only on-site correlations.

Although our fermion sampling method is based on a noninteracting fermionic system, it can still be applied in an interacting case with the reweighting technique.
For example, the expectation of a physical observerable $O$ can be calculated as
\be
\langle O \rangle  = \frac{\langle \psi_J|O|\psi_J \rangle}{\langle \psi_J|\psi_J \rangle}
                = \frac{\sum_x \langle \psi_J|x\rangle \langle x|O|\psi_J \rangle}{\sum_x \langle \psi_J|x\rangle \langle x|\psi_J \rangle}.
\ee
Since the Jastrow factor $\hat{J}$ is diagonal in the position basis, the R.H.S can be transformed as 
\be
\frac{\sum_x J^2(x) |\psi_0(x)|^2 (\langle x|O|\psi_J \rangle / \langle x|\psi_J \rangle)}{\sum_x J^2(x) |\psi_0(x)|^2},
\ee
where $J(x)=\langle x|\hat{J}|x\rangle$ and $\psi_0(x)=\langle x|\phi_0 \rangle $ are both scalar functions.
Then we implement the standard reweighting procedure which is shown as 
\begin{align}
    \langle O \rangle  &= \left(\frac{\sum_x J^2(x) |\psi_0(x)|^2 o_L(x)}{\sum_x |\psi_0(x)|^2}\right) \bigg/ \left(\frac{\sum_x J^2(x) |\psi_0(x)|^2}{\sum_x |\psi_0(x)|^2}\right) \notag \\
                &= \langle J^2 o_L \rangle_0 / \langle J^2 \rangle_0.
\label{eq:reweight}
\end{align}
Here we denote $o_L(x)=\langle x|O|\psi_J \rangle / \langle x|\psi_J \rangle$ and $\langle ... \rangle_0$ labels the expectation value derived from a free fermion ensemble. 
Therefore we can calculate the results for an interacting system by sampling the corresponding noninteracting ensemble and then reweighting according to \ref{eq:reweight}.

\section{Implementation details for text summarization}
\subsection{Feature to magnitude mapping}
The feature to magnitude mapping is achieved by a log-linear model. For the $i$th sentence in a given context $x$, we define a real valued feature function $\boldsymbol{f}(x,i) \in \mathbb{R}^{m}$. This manually assigned function concerns information of the sentence other than its content. The magnitude, is consequently evaluated by 
\begin{equation}
    r(x,i) = \exp(\boldsymbol{\theta} \cdot \boldsymbol{f}(x,i))
\end{equation}
The parameter $\boldsymbol{\theta}$ is a linear weight of each terms in feature function, and is trained with SGD method in our dataset to maximize the overlap between machine summary and human summary. In this specific case, SGD method is guaranteed to find the maximum as our characterization of the overlap can be shown to be concave \cite{ImageDPP_kulesza2011k}.

The features we considered are as follows:
\begin{itemize}
    \item A constant to control the summary length. The expected length of a summary vary from different purpose, and therefore must be controlled manually.
    \item Length of the sentence. Generally speaking, a medium length sentence makes a good summary. Short sentences contains inadequate information and long sentences are difficult to read at first glance. We adopted a polynomial to characterize the quality based on sentence length, including up to the cubic term.
    \item Position in the article. Representative sentences usually appears at the beginning and end of an article. We included the first three powers as well.
    \item Dialogue and personal pronouns. Summary can usually be subjective and narrative. We want to avoid choosing sentences with too much personal pronouns that hinder the information of names, and also avoid objective statements. We included a binary judgement of whether the sentence belongs to a dialogue, and also the number of personal pronouns appears in the sentence.
\end{itemize}

\subsection{MBR decoding}
DPP model produce probabilistic results, so we must excluded summaries with low quality. Here we select a summary result with minimum Bayes risk (MBR) of producing unrelated result\cite{GOEL2000115}. Let $\{s_i\}_{i=1}^N$ be a series of summaries from a text, and $\{\boldsymbol{v}_i\}$ be the corresponding unit vectors of these summaries from word embedding algorithm. The selected summary $s_{MBR}$, according to MBR decoding method, is given by $s_i$ with 
\begin{equation}
    \max_{i} \frac{1}{N}\sum_{j=1}^N \boldsymbol{v}_i\boldsymbol{v}_j
\end{equation}
In the test of our program, 400 individual summaries are taken from each text for MBR decoding. The value was picked to maintain a balance between summary quality and running time. The program generally produce stable results (seldom vary from different trials) at this level.

\section{Fast fermion sampling for continuous models} 
In the main text, we provide an algorithm for a discrete lattice model having computation complexity ${\cal O} (N^2 L)$, with $N$ the fermion particle number and $L$ the number of lattice sites. In application to a continuous model where the fermions' position take continuous values, this algorithm becomes inefficient---we need to discretize the system for which a large number of lattice sites $L$ is typically required to minimize the discretization error. Here we describe how our fast fermion sampling algorithm adapts to continuous models. In the first step, 
we still generate a vector ${\bf m}$ as a random permutation of ${\bf n}$ ({\it see main text}). Then in generating the fermion position $x_k$ from the conditional probability distribution $P(x_k| x_1, \ldots x_{k-1}; {\bf m})$, we perform Markov chain sampling instead of the explicit sampling as used for discrete models.  A trial random continuous update is proposed ($x_k \to x_k + \delta x$), and  then accepted with probability 
\be 
P_{\rm AC} (x_k \to x_k + \delta x) = \frac{|{\rm det} [U_{(x_1 \ldots x_k+\delta x), (m_1 \ldots m_k)} ]|^2 }{|{\rm det } [U_{(x_1 \ldots x_k), (m_1 \ldots m_k)} ] |^2}
= \frac{|{\bf u}_{x_k+\delta x} ^T {\bf h} ( {\bf u}_{x_1}, {\bf u}_{x_2}, \ldots, {\bf u}_{x_{k-1}})|^2}{|{\bf u}_{x_k} ^T {\bf h} ( {\bf u}_{x_1}, {\bf u}_{x_2}, \ldots, {\bf u}_{x_{k-1}})|^2}, 
\ee 
where \{${\bf u}_{x_1}, {\bf u}_{x_2}, \ldots, {\bf u}_{x_{k}}$\} are $k$-dimensional row vectors in $U_{(x_{1} \ldots x_{k}), (m_1 \ldots m_k)}$ and ${\bf h}$ is a $k$-dimensional normal vector perpendicular to the $k-1$ vectors \{${\bf u}_{x_1}, {\bf u}_{x_2}, \ldots, {\bf u}_{x_{k-1}}$\}.
The update is iterated for $M_{\rm cond}$ times in order to let the Markov chain equilibrate. This is the only modification we need to make  for sampling continuous models to the fast fermion sampling algorithm. The sampling complexity is then ${\cal O} (N^3)$, which is independent of $L$. 

We emphasize that  Markov chain is used here to sample a single-variable probability distribution. The autocorrelation length in this sampling is at most a constant, which does not scale up with the system size or the fermion number. With $M_{\rm cond}$ chosen to be larger than the autocorrelation length, the entire fermion sampling algorithm should still be taken as an explicit sampling algorithm, free of any diverging autocorrelation length problem. 

It is worth mentioning here that the sampling algorithm for continuous models described above also applies to problems with large $L$. We simply need to replace the continuous update $x_k \to x_k + \delta x$ by a discrete update $x_k \to x_k'$. The sampling complexity is also ${\cal O} (N^3 )$.

\end{widetext}

\end{document}